\input{aipcheck}

\documentclass[
    ,final            
  ]
  {aipproc}

\layoutstyle{6x9}

\begin{document}

\title{Testing evolutionary tracks of Pre-Main Sequence stars: the case of
HD113449}


\author{F. Cusano}{
  address={Thüringer Landessternwarte Tautenburg,  
 Sternwarte 5, D - 07778 Tautenburg}
 }

\author{E. W. Guenther}{
 address={Thüringer Landessternwarte Tautenburg,  
 Sternwarte 5, D - 07778 Tautenburg}
 }

\author{M. Esposito}{ 
address={Hamburger Sternwarte,
 Gojenbergsweg 112,
 21029 Hamburg}
}
\author{M. Mundt}{
address={Max-Planck-Institut für Astronomie,
 Königstuhl 17, 
 D - 69117 Heidelberg,
 Germany}
 }
\author{E. Covino }{
address={Osservatorio Astronomico di Capodimonte,
Salita Moiariello 16,   
 80131 - Napoli    }
 }
\author{ J. M. Alcal\`{a}}{
address={Osservatorio Astronomico di Capodimonte,
Salita Moiariello 16,   
 80131 - Napoli    }
 }

\begin{abstract}
 Evolutionary tracks are of key importance for the understanding of
 star formation. Unfortunately, tracks published by various groups
 differ so that it is fundamental to have observational tests. In
 order to do this, we intend to measure the masses of the two
 components of the Pre-Main Sequence (PMS) binary HD113449 by
 combining radial velocity (RV) measurements taken with HARPS, with
 infrared interferometric data using AMBER on the VLTI. The
 spectroscopic orbit that has already been determined, combined with
 the first AMBER measurement, allows us to obtain a very first
 estimation of the inclination of the binary system and from this the
 masses of the two stars. More AMBER measurements of HD 113449 are
 needed to improve the precision on the masses: in the ESO period P82
 two new measurements are scheduled.
\end{abstract}

\maketitle


\section{Introduction}

  The most fundamental parameter of a star is its mass which
  determines almost everything about its birth, life and death. For PMS
  stars this parameter is practically always derived by comparing the
  location of the star in the Hertzsprung-Russell Diagram (HRD) with theoretically calculated
  evolutionary tracks. Unfortunately, the evolutionary tracks
  published by various authors differ considerably due to the
  differences in the input
  physics \cite[e.g.][]{dantona94,baraffe98,palla99}.  In order to
  test and calibrate the tracks it is necessary to determine the
  masses for a number of young stars. One possibility is to combine RV
   data and interferometric data in order to
  measure the masses of young stars in spectroscopic binary (SB)
  systems. In 1999, we thus initiated a spectroscopic survey for PMS
  binaries and detected 15 systems suitable for AMBER
  (see \cite{petrov07} for a description of the instrument)
  observations \cite{guenther07}. Unfortunately, all of these stars are
  too faint for AMBER at the present stage. In addition to these stars
  we found a young SB1 system in the course of an RV-survey with HARPS:
  HD113449. HD113449 is at a distance of only 22.1$\pm$0.6 pc
  (\cite{perryman97}). The primary has a spectral type of
  G5V, and an equivalent width of the LiI6708$\lambda$ line of
  160m\AA. The secondary is presumably a late K or early M star. The
  binary thus consists of two post T Tauri stars, located in a zone of
  the HRD where the radiative core is
  supposed to form. This region of the HRD is of particular interest
  because the only stars where the masses have been determined are the
  two components of the binary system HD98800B (\cite{boden05}), and
  these only match the evolutionary tracks if either a very low
  metallicity is assumed, or if the radiative core forms at an earlier
  stage than is presently assumed.

\section{Observations and Data Reduction}

Using HARPS, we have observed HD113449 for 2.5 years. The HARPS
spectra cover the wavelength region from 378 to 691 nm at
R=115000. The HARPS pipeline was used in order to measure the RV of
this star.  We henceforth solved the SB1-orbit (Fig. 1), determining
the  orbital parameters given in Tab. 1. The inclination of the orbit
of the binary with respect to the line of sight and the longitude of
the ascending node $\Omega$ can not be derived from the
RV-measurements. For this, the binary system has been observed also
with AMBER, the near-infrared spectro-interferometric VLTI
instrument. Observations have been performed the night between the
20-21st of March 2008.  AMBER was used in Low-Resolution mode
combining the light coming from the UT2, UT3 and UT4. In the same
night, shortly before HD113449, a standard star, HD111998, of known
diameter ($\theta\sim0.30$mas) has been observed to calibrate the
visibility.  The data have been reduced at ESO using the AMBER-pipeline. The
reduced data have been converted to ascii format running an automatic
procedure in the ESO-pipeline and a set of procedures specifically written by us
have
been used to calibrate the square visibility.  In the first step we
considered the calibrator to be unresolved on all the three baselines.
This assumption leads to a small error on the calibrated visibility,
especially for the largest baseline (UT1-UT3=77.9m). The larger
scatter of the data on the right of Fig. 2 is presumably related to
this assumption. As a next step we will correct for this. For HD113449 also an 
IR high-resolution spectra (R=100000,2236-2293nm) has been taken with CRIRES. Standard
procedure have been used to reduce it.

\begin{figure}
  \includegraphics[width=120mm,height=70mm]{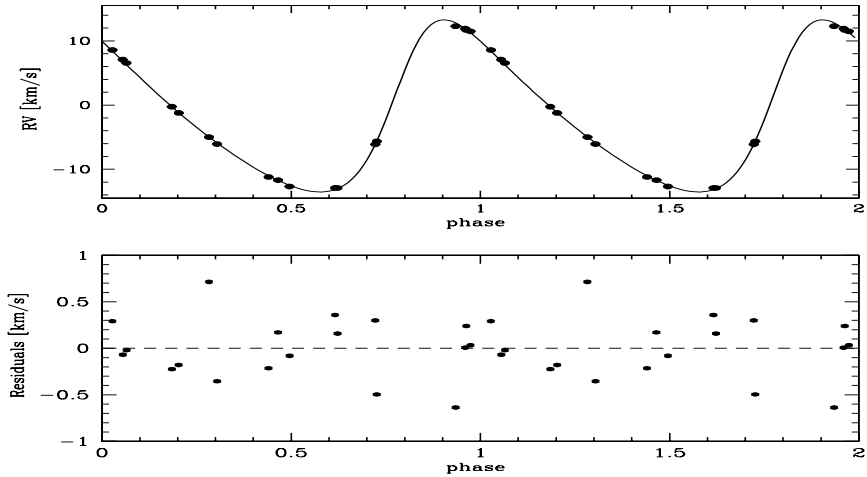}
  \caption{The phase-folded RV-measurements of HD113449. The orbital period is 216 days.}
\end{figure}

\begin{table}
\footnotesize
\begin{tabular}{c c}

\hline
element &value \\
   \\
\hline
P & $215.9\pm0.1$ d\\
T$_0$[HJD]&$2453411\pm1$ \\
$\gamma$& $-1.79\pm0.02$ km s$^{-1}$\\
K$_1$  & $13.40\pm0.02$ km s$^{-1}$\\
e & $0.300\pm0.005$\\
$\omega$ & $114.6\pm0.5^{\rm{o}}$ \\
a$_1$sin i & $0.254\pm0.001$ AU\\
f(m) & $0.0467\pm0.0006$ M$_\odot$\\
\hline
\end{tabular}
\caption{Orbital elements of HD113449}
\label{tab:a}
\end{table}

\section{Analysis and Results}

AMBER in the LR configuration produces spectral dispersed fringes
along the J, H and K bands. With a brightness of J=6.0, H=5.7, and
K=5.5, HD113449 is close to the limit of AMBER. The S/N in J, and H
data is in fact so low that the data can not be used and our analysis thus
is based on the K-band.  To derive a first estimation of the
inclination of the orbit and of the ascending node, $\Omega$, we
compared the observations with a set of theoretical models. For this
aim we wrote a software that simulates the visibility of a binary
system given all the orbital parameters, the baselines and the epoch
of observation. Given the distance to the system and the predicted
radius both the stars are point sources.  With this assumption the
formula used for the visibility is simply: $V^2 = (1 +
r^2+2rcos({\bf{\rho}} {\bf{B}}/\lambda))/(1+r)^2$, where $r$ is the flux
ratio between the two components, {$\bf{\rho}$} the angular
separation, ${\bf{B}}$ the projected baseline and $\lambda$ the
wavelength of observation.  Fixing the parameters that were obtained
from the RV-data, we vary just i and $\Omega$, fixing also the flux in
the K-band of the two stars assuming their spectral type. Considering
that the primary is a G5V and the secondary an early M-star the flux
ratio in the K-band is: $r$=$0.21\pm0.02$. 
 Using a
$\chi^2$ analysis we derive a value for the inclination of
i=$63\pm3^{\rm{o}}$ and for the longitude of the ascending node
$\Omega$=$102\pm8^{\rm{o}}$. Fig. 2 shows the calibrated visibility
squared vs. radius (in unit of Baseline/wavelength) of HD113449 during
45 minutes of observations. Also shown are the modeled visibility that
best match the observations, from which we derive the inclination and
the $\Omega$. 
From a first analysis of the CRIRES spectra (Fig. 3)
we derive also a mass-ratio of primary to the secondary of $0.57\pm0.05$.
Combining the value found for the inclination, the mass
function and the mass ratio we estimate the
masses to be M$_1=0.88\pm0.13$M$_\odot$ and
M$_2=0.50\pm0.07$M$_\odot$ for the primary and the secondary, respectively.  More observations with
AMBER are needed to improve the derived orbital parameters. The AMBER 
and CRIRES data are also still being analysed. 

\begin{figure}
  \includegraphics[width=120mm,height=70mm]{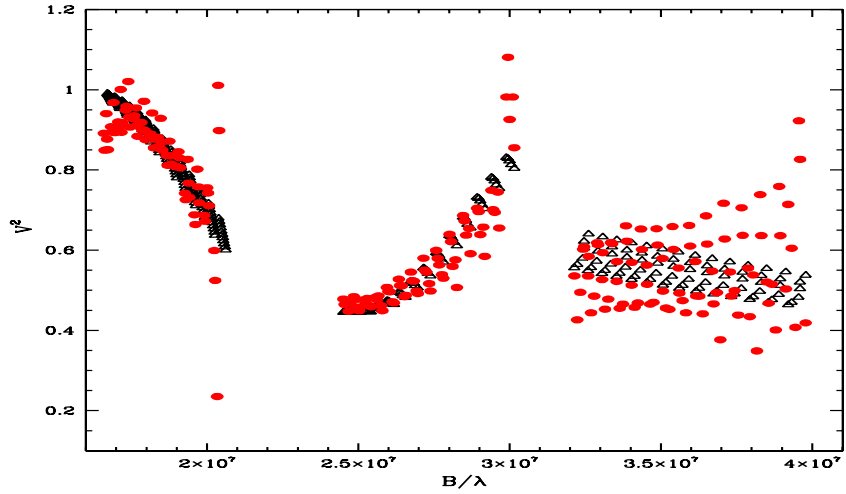}
  \caption{The red dots are the calibrated squared visibility of HD113449.
   Shown as black open triangles are models with i$\sim$ 63$°$ and 
   $\Omega\sim102°$ which match the observations best.
}
\end{figure}

\begin{figure}
  \includegraphics[width=120mm,height=75mm]{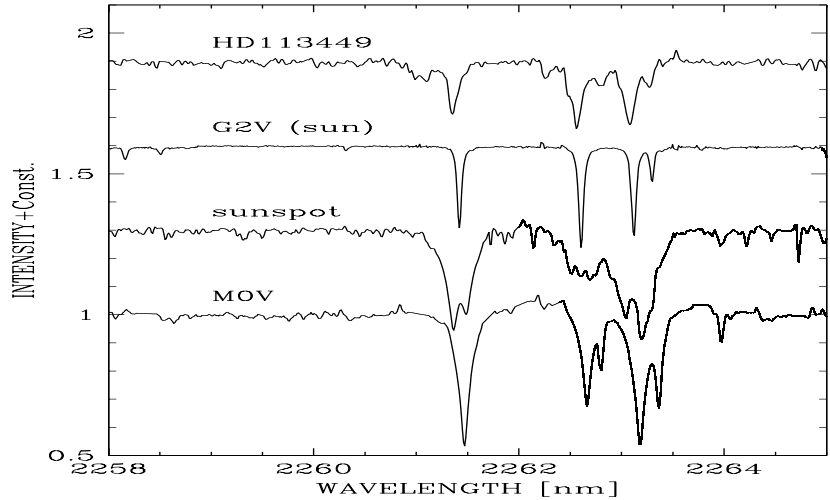}
  \caption{From top to bottom: a small part of the CRIRES spectrum of HD113449, FTS
   spectrum of the solar photosphere and a sunspot (M3V), and finally
   CRIRES spectrum of the M0V star GJ9847. Clearly visible
   in the spectrum of HD113449 are the lines of the G5V primary,
   and slightly blue-shifted to these the lines of the secondary.
   The lines in the spectrum of the sunspot (as well as GJ9847) 
   are splitted into the Zeeman components. 
}
\end{figure}

\end{document}